\documentclass[twocolumn]{aastex631}
\usepackage{amsmath}
\usepackage{float}
\begin{document}

\title{Water and an escaping helium tail detected in the hazy and methane-depleted atmosphere of HAT-P-18b from JWST NIRISS/SOSS}

\author[0000-0002-3263-2251]{Guangwei Fu}
\affiliation{Department of Physics and Astronomy, Johns Hopkins University, Baltimore, MD 21218, USA; guangweifu@gmail.com}

\author[0000-0001-9513-1449]{Néstor Espinoza}
\affiliation{Space Telescope Science Institute, 3700 San Martin Drive,
Baltimore, MD 21218, USA}

\author[0000-0001-6050-7645]{David K. Sing}
\affiliation{Department of Physics and Astronomy, Johns Hopkins University, Baltimore, MD 21218, USA}
\affiliation{Department of Earth \& Planetary Sciences, Johns Hopkins University, Baltimore, MD, USA}

\author[0000-0003-3667-8633]{Joshua D. Lothringer}
\affiliation{Department of Physics, Utah Valley University, Orem, UT 84058, USA}

\author[0000-0002-2248-3838]{Leonardo A. Dos Santos}
\affiliation{Space Telescope Science Institute, 3700 San Martin Drive, Baltimore, MD 21218, USA}

\author[0000-0001-6050-7645]{Zafar Rustamkulov}
\affiliation{Department of Earth \& Planetary Sciences, Johns Hopkins University, Baltimore, MD, USA}

\author[0000-0001-5727-4094]{Drake Deming}
\author[0000-0002-1337-9051]{Eliza M.-R. Kempton}
\author[0000-0002-9258-5311]{Thaddeus D.\ Komacek}
\affiliation{Department of Astronomy, University of Maryland, College Park, MD 20742, USA; guangweifu@gmail.com}

\author[0000-0002-5375-4725]{Heather A. Knutson}
\affiliation{Division of Geological and Planetary Sciences, California Institute of Technology, 1200 East California Boulevard, Pasadena, CA 91125, USA}

\author[0000-0003-0475-9375]{Loïc Albert}
\affiliation{Institut de Recherche sur les Exoplan\`etes (iREx), Universit\'e de Montr\'eal, D\'epartement de Physique, C.P. 6128 Succ. Centre-ville, Montr\'eal, \\ QC H3C 3J7, Canada.}

\author{Klaus Pontoppidan}
\author{Kevin Volk}
\author{Joseph Filippazzo}
\affiliation{Space Telescope Science Institute, 3700 San Martin Drive, Baltimore, MD 21218, USA}

\begin{abstract}

JWST is here. The early release observation program (ERO) provides us with the first look at the scientific data and the spectral capabilities. One of the targets from ERO is HAT-P-18b, an inflated Saturn-mass planet with an equilibrium temperature of $\sim$850K. We present the NIRISS/SOSS transmission spectrum of HAT-P-18b from 0.6 to 2.8$\mu m$ and reveal the planet in the infrared beyond 1.6$\mu m$ for the first time. From the spectrum, we see clear water and escaping helium tail features in an otherwise very hazy atmosphere. Our free chemistry retrievals with ATMO show moderate Bayesian evidence (3.79) supporting the presence of methane, but the spectrum does not display any clearly identifiable methane absorption features. The retrieved methane abundance is $\sim$2 orders of magnitude lower than that of solar composition. The methane-depleted atmosphere strongly rejects simple equilibrium chemistry forward models with solar metallicity and C/O ratio and disfavors high metallicity (100 times) and low C/O ratio (0.3).
This calls for additional physical processes such as vertical mixing and photochemistry which can remove methane from the atmosphere.

\end{abstract}
\nopagebreak

\section{Introduction}

The era of JWST has finally arrived \citep{the_jwst_transiting_exoplanet_community_early_release_science_team_identification_2022}. The unparalleled infrared capability and photometric precision from JWST will evermore change our understanding of exoplanet atmospheres. With the expanded spectral windows into the infrared, we can now identify and measure new molecular features that have never been studied in detail before. Methane, an abundant chemical species on Earth and many solar system planets, has yet to be detected in a short-period exoplanet atmosphere. It is one of the most significant byproducts of life activities on Earth, and detecting methane is one of the many required steps to eventually finding evidence of carbon-based life as we know it beyond Earth. 

Methane is ample in solar system gas giants (Jupiter, Saturn, Uranus, and Neptune) where the upper atmosphere is cooler than 200K. The lack of evidence for methane in short-period exoplanet atmospheres has been puzzling \citep{kreidberg_absence_2019, stevenson_possible_2010} as equilibrium chemistry forward models predict the rise the methane abundance in sub-1000K exoplanet atmospheres \citep{moses_disequilibrium_2011, benneke_sub-neptune_2019}. Methane can be destroyed at high temperatures \citep{zahnle_methane_2014} and it also has a short photochemical lifetime \citep{kempton_atmospheric_2012, kasting_photochemistry_1983, thompson_case_2022}. Measurements of methane abundances in exoplanet atmospheres will help us to better understand atmospheric vertical mixing based on interior thermal structure and photochemistry\citep{fortney_beyond_2020}.

HAT-P-18b is a Saturn-like planet with a mass of 0.197M$_{Jup}$ and a radius of 0.995R$_{Jup}$ \citep{hartman_hat-p-39bhat-p-41b_2012}. It has an equilibrium temperature of $\sim$850K. The relatively low temperature compared to other hot Jupiters ($>$1000K) make it an intriguing target for atmospheric characterization to study disequilibrium chemistry \citep{fortney_beyond_2020}, especially its impacts on methane abundance \citep{tsai_comparative_2021, zahnle_methane_2014}.

Here we present the transmission spectrum of HAT-P-18b obtained using JWST Near Infrared Imager and Slitless Spectrograph single object slitless spectroscopy (NIRISS/SOSS), which covers $\sim$0.6 to 2.8 $\mu $m. Our data are sensitive to multiple molecular features from water and methane which should allow us to robustly detect and measure their abundance. The long continuous wavelength coverage extending into the optical also allows us to place tight constraints on HAT-P-18b's aerosol properties. In this Letter, we report the clear detection of water, an escaping helium tail, a sub-Rayleigh haze scattering slope, and methane depletion in the atmosphere of HAT-P-18b.

\section{Observations and data reduction}

This dataset was collected as part of the ERO program \citep[Program ID: 2734;][]{pontoppidan_jwst_2022}. A transit of HAT-P-18b was observed using NIRISS/SOSS on June 13th 2022 with the GR700XD grism and the CLEAR filter. The SUBSTRIP256 detector was used to capture both order 1 and order 2 of the dispersed spectra. A total of 7.15 hours of time series observation (TSO) was taken to ensure sufficient out-of-transit baseline coverage. There are 9 groups per integration with each integration taking 54.94 seconds and a total of 469 integrations. 

\begin{figure*}
\centering
  \includegraphics[width=\textwidth,keepaspectratio]{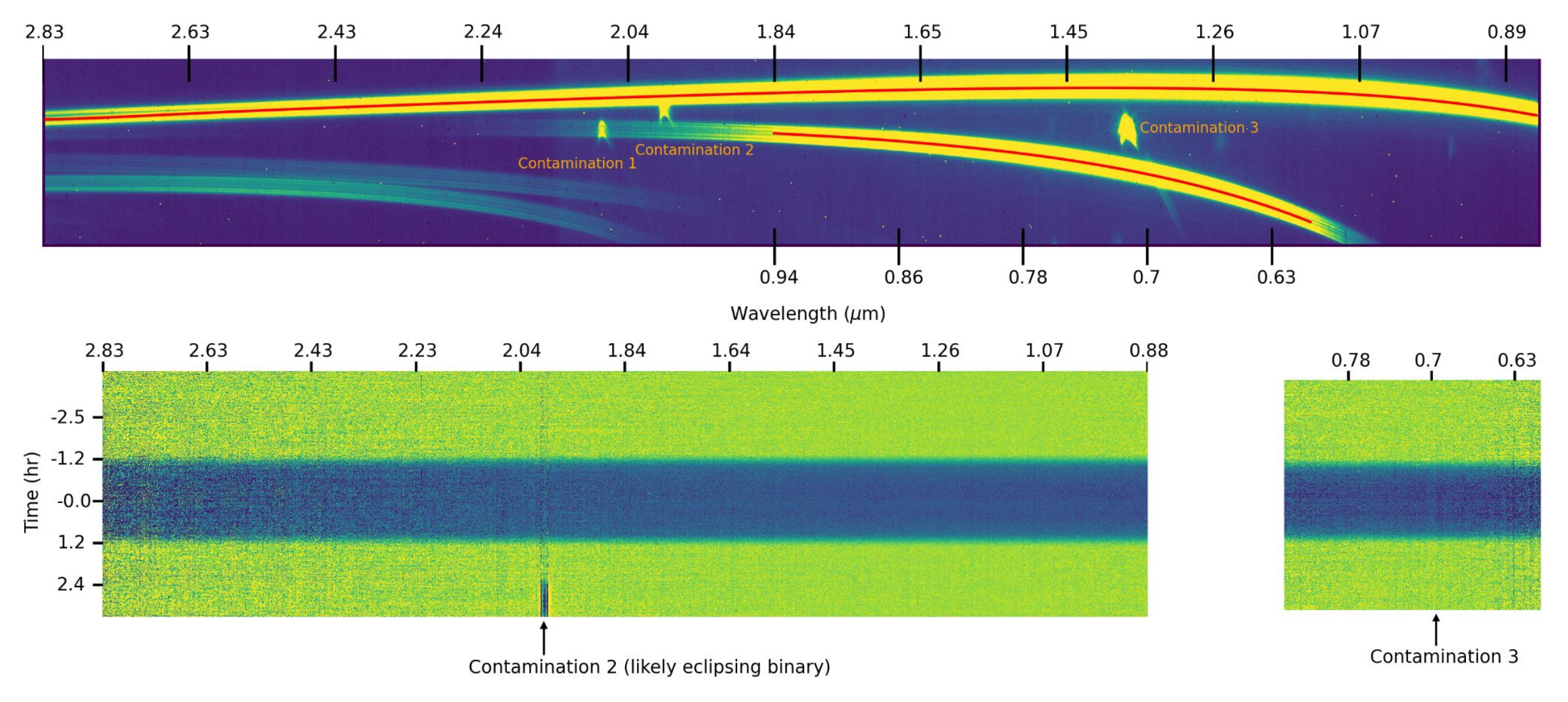}
  \caption{The top panel shows one of the integrations from \texttt{rateints.fits} files, the red lines show the first and second order spectra traces. The wavelength-dependent zodiacal light background and three zeroth order nearby contamination stars are visible in the image. The bottom panel shows the reduced light curve at every pixel bin. Contamination 1 has a minimal effect on the planet spectrum as it is far away from the first order and not extracted in the second order. We detect a time-varying flux level in contamination 2 ($\sim$12$\%$ dimming) which is likely caused by an eclipsing binary. Contamination 3 has a spike going across the second order spectrum which leads to wavelength-dependent dilution effects. We removed the contaminated parts of the spectrum from  sources 2 and 3 in the analysis.}
  \label{fig:fig1}
\end{figure*}

\subsection{Data reduction}

We begin our data reduction with the \texttt{rateints.fits} countrate data downloaded from MAST archive on July 28th 2022. The countrate products are generated from the JWST Science Calibration Pipeline stage 2a which fits the accumulating signal ramp for every pixel in each integration. One integration of the \texttt{rateints.fits} file is shown in the top panel of Figure \ref{fig:fig1}. The image from the SUBSTRIP256 subarray has a dimension of 2048 $\times$ 256 pixels with wavelength increasing from right to left. Since SOSS mode is slitless, contamination from nearby stars and the sky background can dilute transit signals. The first step is to remove the zodiacal light background which increases with wavelength with a sharp drop around 2.1$\mu m$ caused by the edges of the pick-off mirror. We used the smoothed background template from the commissioning observation (Observation 5 of Program 1541) downloaded from \texttt{jwst-docs}\footnote{https://jwst-docs.stsci.edu/}. The background template is first scaled based on the median of the bottom 20 percentile pixels between columns 300 and 500 ($\sim$2.3 to 2.5 $\mu m$) and then subtracted from every integration. Since we do not expect the zodiacal light background to be time-dependent on our 7.15 hours timescale, we subtract the same scaled background template from all integrations. 

Next, we determine the first and second-order spectral trace locations by first taking the PSF from the bright first-order spectral region ($\sim$1.5 $\mu m$) and then cross-correlating it at every column. The peak cross-correlation values at each column as a function of column pixel values are then fitted with a polynomial to smooth out the effects of contamination and bright outlier pixels. The first-order spectrum is then extracted with a width of 60 pixels centered at the trace as shown in red lines in the upper panel of Figure \ref{fig:fig1}. The \texttt{scipy.interpolate.interp1d} function is used at each column to interpolate each extraction at the sub-pixel levels. The same steps are then used to extract the second-order spectra after the first order is extracted from the 2D image. 

After the first and second-order spectra have been extracted and aligned in the y-axis, we applied a median filter along each row to remove all outlier bright pixels within the 60-pixels wide spectral cut. Then we subtract the median of the combined top and bottom 15 pixels at each 60-pixels wide column to remove the 1/f noise. At last, we sum the middle 30 pixels at each extracted spectral column for every integration to get the light curves for all wavelength channels. All the cleaned light curves at each column are shown in the lower panel of Figure \ref{fig:fig1}.

\subsection{Light curve fitting}

\begin{figure*}
\centering
  \includegraphics[width=\textwidth,keepaspectratio]{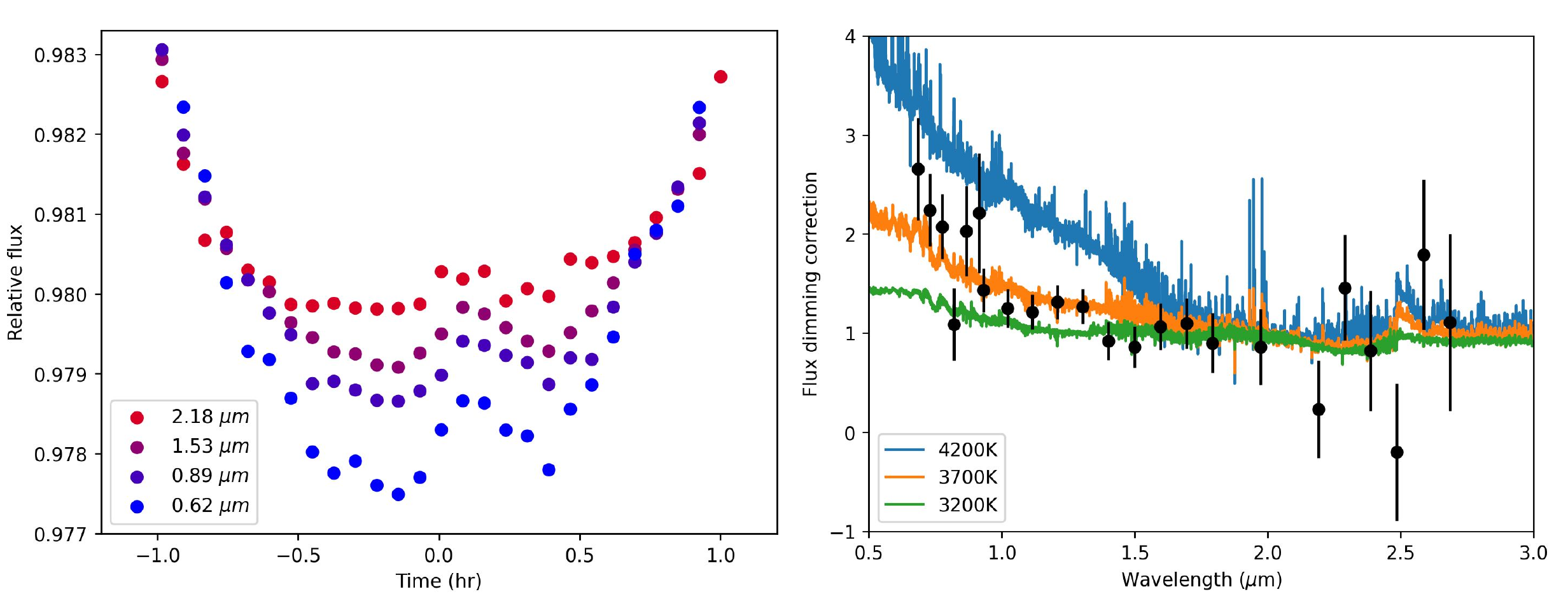}
  \caption{The transit light curve has a clear starspot crossing event starting around mid-transit (left panel). The starspot crossing affects the light curve as a function of wavelength with shorter wavelengths being the most affected. We first measure the amplitude of the starspot at each wavelength and then use the PHOENIX stellar models to obtain a best-fit starspot temperature of 3596$\pm$31K (right panel). Assuming other unocculted starspots are at a similar temperature, we then use the stellar flux variability level seen in TESS observations to estimate the dilution effect. We obtain a $<$20ppm level dilution effect between 0.6 and 2.8$\mu m$ which is much smaller than the uncertainties in the transmission spectrum.}
  \label{fig:fig2}
\end{figure*}

We first fit the white light data using the \texttt{Batman} \citep{kreidberg_batman_2015} transiting light curve model with \texttt{emcee} \citep{foreman-mackey_emcee_2013}. There are a total of six free parameters in the fit including the mid-transit time, transit depth, semi-major axis, inclination, linear slope, and a constant offset. There is a starspot crossing event during the transit which changes the transit light curve shape in a wavelength-dependently way (left panel of Figure \ref{fig:fig2}). We mask out the starspot crossing (integration 240 to 272) during the whitelight fit. After subtracting the best-fit model, we fit a polynomial model to the starspot crossing residual. Since the starspot shape varies as a function of wavelength due to its different temperature relative to the stellar photosphere, we use the best-fit model from the white light fits and scale it for each wavelength \citep{sing_hubble_2011}.

The best-fit whitelight mid-transit time (2459743.853395$\pm$ 0.000023 MJD), semi-major axis (a/R$_{star}$=16.52$\pm$0.06), and inclination (88.66$\pm$0.03 degrees) are then fixed to fit the light curve from each column. There are four free parameters in the fit including the transit depth, a linear slope, a constant offset, and a scaling factor for the starspot crossing model. We used the \texttt{scipy.optimize.curvefit} function for individual wavelength channel fits. For the limb darkening, we used quadratic coefficients from the 3D
stellar model in the Stagger-grid \citep{magic_stagger-grid_2015}. For the wavelength solution, we used the \texttt{jwst-niriss-wavemap-0021.fits} wavelength map file from the JWST Calibration Reference Data System (CRDS).

We compared our final reduced transmission spectrum to an independent reduction from Néstor Espinoza using the pipeline \texttt{TransitSpectroscopy}\footnote{https://github.com/nespinoza/transitspectroscopy} and they are in excellent agreement as shown in the top panel of Figure \ref{fig:spectrum}.

\begin{figure}
\centering
  \includegraphics[width=0.45\textwidth,keepaspectratio]{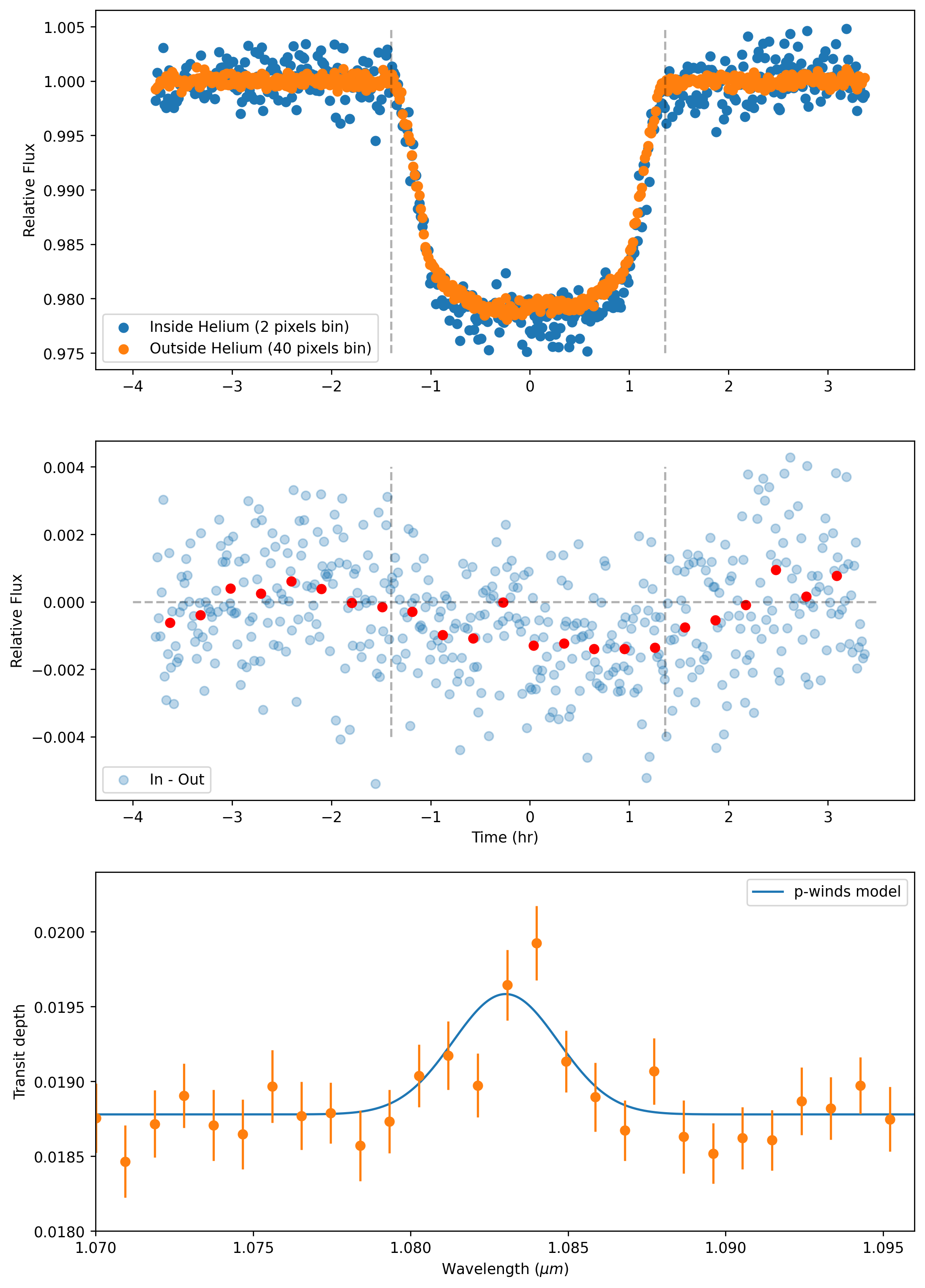}
  \caption{We detect escaping helium in HAT-P-18b. The top panel shows the light curve centered at the 1083 nm helium line versus whitelight light curve from nearby pixels. The middle panel shows the residual (binned in red) from subtracting the outside helium light curve from the inside helium light curve. The residual clearly shows a deeper transit and a tail after the egress. The vertical grey dash lines show the first and fourth contact and the horizontal grey dash line is the zero residual level. The bottom panel is the transmission spectrum of HAT-P-18b at pixel level overplotted with the best-fit {\tt p-winds} model.}
  \label{fig:helium}
\end{figure}

\subsection{Contamination and dilution}
To ensure the accuracy of our transmission spectrum we have investigated possible sources of contamination and dilution. We identify the following three sources that could potentially affect the planet spectrum.

\subsubsection{Zeroth order contamination}

As NIRISS/SOSS is a slitless spectrograph, nearby stars can overlap with the target's spectral trace. There are three clear zeroth order nearby star contamination in the observation as shown in the upper panel of Figure \ref{fig:fig1}. Contamination 1 is on the second order spectral trace but fortunately, it only affects the longer wavelength portion of the second order which is not extracted due to the relatively low counts. It is also too far away from the first-order spectrum to cause dilution. Contamination 2 is more complex as it sits directly on the first order dispersion and its brightness also varies in time (Figure \ref{fig:fig1}). It is likely an eclipsing binary with ingress starting $\sim$2.4 hours after HAT-P-18b's mid-transit, the observation only catches a partial egress but it has a V-shaped transit with a maximum dimming of $\sim$12$\%$. We masked this part of the spectrum (1.98 - 2.01$\mu m$) in the analysis. Contamination 3 is located between the first and second order, but it has a long tail going across the second order around 0.7$\mu m$. The tail leads to a wavelength-dependent dilution in the transit depth, therefore we removed the 0.68 to 0.72$\mu m$ part of the spectrum in the analysis. 

\subsubsection{A nearby source from GAIA}

There is a nearby star (2.66$\arcsec$ away) in the GAIA database \citep{gaia_collaboration_gaia_2018} (source id: 1334573817793501696) which could cause dilution in the overlapped spectral dispersion. The companion has a delta magnitude of 4.587 and 4.898 fainter than HAT-P-18b respectively in the GAIA blue and red passbands. We performed a GAIA colors fit of the companion relative to HAT-P-18 (T$_{eff}=4800K$) and obtained a best-fit T$_{eff}$ of 5500K$\pm$300K. Then we normalize the 5500K and 4800K PHOENIX models to the delta magnitude and calculate the dilution effect on the transit depth. Assuming the worst-case scenario of complete companion spectral overlap, the resulting dilution is less than 20ppm between the 0.6 and 3 $\mu m$ regions of the transit spectrum. This is much smaller than the uncertainties on the final transmission spectrum, and we only expect slight spectral overlap from the companion base depending on the position angle of the instrument. Therefore, we conclude it is safe to ignore this GAIA nearby source. 
We also checked archival infrared high-contrast imaging data and there are no other known nearby sources \citep{ngo_friends_2015, ngo_friends_2016}.

\subsubsection{Unocculted starspots}

Unocculted starspots can lead to a wavelength-dependent dilution effect on the transmission spectrum \citep{agol_climate_2010, pont_prevalence_2013}. To estimate the magnitude of this effect, we follow the steps detailed in section 3.5 of \cite{sing_hubble_2011}. We first fit for the starspot temperature using the best-fit starspot scaling factor at each wavelength. The best-fit starspot temperature is 3596$\pm$31K assuming a host star temperature of 4800K (Figure \ref{fig:fig2}). Next, we estimate the photometric stellar variability amplitude from TESS observations which observed the star during sectors 25 and 26. After removing the transits in TESS, we used a Lomb-Scargle periodogram to find a best-fit variability period of $\sim$14.3 days which is consistent with the reported stellar rotational period of 14.66$\pm$0.03 days \citep{everett_spectroscopy_2013}. The corresponding variability amplitude is 0.0724$\%$ and it is within the 0.1$\%$ level mean R-band brightness variation observed from the 12-month ground-based monitoring campaign \citep{kirk_rayleigh_2017}. Finally, we use equations (4) and (5) from \cite{sing_hubble_2011} to calculate an expected dilution effect of $<$20ppm between 0.6 and 3 $\mu m$ in the transmission spectrum. Since this effect is much smaller than the uncertainties on the transmission spectrum, we conclude it is safe to ignore the unocculated starspots. 

\section{Detection of excess helium absorption and tail}

We detect excess metastable 1083 nm helium absorption in the transmission spectrum. The helium line is not fully spectrally resolved with the SOSS (R$\sim$550 at 1083 nm) but the transit depth from a 2 pixels wide bin (one resolution element) centered at 1083.065 and 1083.998 nm is visibly deeper than that in nearby pixels (Figure \ref{fig:helium}). We measure a helium transit depth of 1.97$\pm$0.016$\%$ with the 2-pixel bin relative to a nearby whitelight (40-pixel bin) depth of 1.88$\pm$0.004$\%$. This is consistent within one sigma to the 2.46$\pm$0.15$\%$ helium transit depth reported in \cite{vissapragada_upper_2022, paragas_metastable_2021} after accounting for the spectral resolution difference. Although our measurement is more diluted than that of \citep{vissapragada_upper_2022}, our detection of excess absorption is more statistically significant (5.6 sigma vs 4.3 sigma). 

We also see evidence of a helium tail at egress (Figure \ref{fig:helium}). We subtracted the nearby whitelight from the helium light curve and the residual clearly shows asymmetry between ingress and egress. Future detailed follow-up 3D hydrodynamical simulations of the tail may tell us more about the escaping material and stellar environment of HAT-P-18b. A planetary tail from escaping helium has only previously been detected on WASP-107b \citep{allart_high-resolution_2019, spake_post-transit_2021, kirk_confirmation_2020}, WASP-69b \citep{nortmann_ground-based_2018} and HD 189733b \citep{guilluy_gaps_2020}. Previous ground-based observations of HAT-P-18b did not see the tail because the signal-to-noise of their excess absorption detection was lower than that of the detection we report here. This demonstrates the unmatched photometric precision from JWST compared to typical ground-based observatories. Although lower in spectral resolution compared to  ground-based high-resolution spectrographs (R$>$10000), JWST NIRSpec/G140H (R$\sim$2700) is a promising instrument mode for helium observations thanks to its superior stability and precision.

\subsection{Escaping helium modeling}
We fit the metastable helium \citep{seager_theoretical_2000} transmission spectrum between 1.07 and 1.10\,$\mu$m with a one-dimensional, isothermal Parker-wind model \citep{parker_dynamics_1958} based on the {\tt p-winds} code \citep[version 1.3.0;][]{dos_santos_p-winds_2022} to determine the mass loss rate of HAT-P-18b. This code applies the original formulation from \cite{oklopcic_new_2018} along with modeling improvements proposed in \cite{lampon_modelling_2021} and \cite{vissapragada_upper_2022}. We assume the planetary outflow has a He/H fraction of 10/90. We further assume that the high-energy SED of HAT-P-18 is similar to HD~85512, and adopt the MUSCLES spectrum \citep{youngblood_muscles_2016, p_loyd_muscles_2016} of the latter as a proxy when calculating the ionization balance in the planetary outflow. We took into account the Roche lobe effects and the fact that the planet is only irradiated over $\pi$ steradians rather than the $4\pi$ steradians assumed by {\tt p-winds} models. 

Our forward models indicate that the mass loss rate of $3.7 \times 10^{10}$~g\,s$^{-1}$ estimated by \citet{vissapragada_upper_2022} for an outflow temperature of $8000$~K is consistent with the observed JWST transmission spectrum in the metastable He feature (see Fig. \ref{fig:helium}). Since the absorption profile is not resolved by NIRISS, there is a degeneracy between the escape rate and outflow temperature when fitting isothermal Parker-wind models to the data. Because of this degeneracy, solutions with lower mass loss rates will fit the observation just as well if they have lower temperatures. This effect is also seen in the narrow-band photometry technique used by \citet{vissapragada_upper_2022}, and it is likely going to be a limitation in future NIRISS observations of He in other planets as well. High-resolution spectroscopy from the ground can resolve the planetary absorption and break this degeneracy.

\begin{figure*}[h]
\centering
  \includegraphics[width=0.8\textwidth,keepaspectratio]{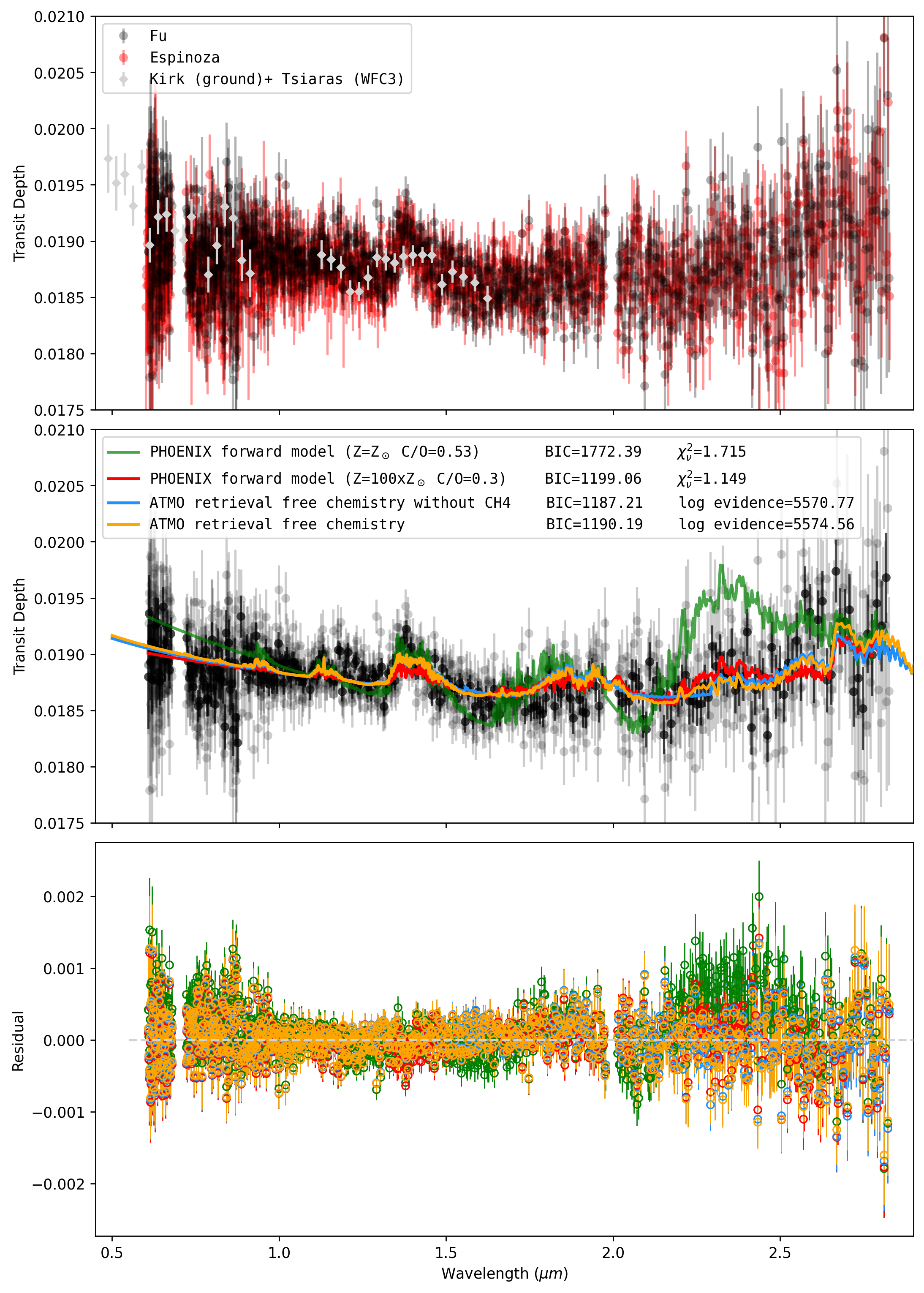}
  \caption{The top panel shows the comparison between two independent data reductions and previous observations from the ground \citep{kirk_rayleigh_2017} and HST/WFC3 \citep{tsiaras_population_2018} with no offsets applied. All data show excellent consistency with each other. The middle panel shows the spectrum (Fu version) overplotted with two PHOENIX forward models and two ATMO free chemistry retrieval best-fit models. The higher R=700 spectrum is in gray and the lower R=200 spectrum is in black. The retrieval is performed on the R=700 spectrum. The lack of clearly identifiable methane feature rules out the solar composition atmosphere. The 100 times solar metallicity and 0.3 C/O forward model provide a decent fit but under predicts the 1.4$\mu m$ water feature while over-predicting the 2.3$\mu m$ methane feature. The ATMO free chemistry retrieval (with H$_2$O, CO$_2$, CO, and CH$_4$) provides the best-fit model, and an identical retrieval without CH$_4$ gives a comparable fit. The log Bayesian evidence difference of 3.79 corresponds to moderate evidence supporting the detection of methane. The BIC metric slightly favors the retrieval without methane due to one less free fitting parameter. The bottom panel shows the corresponding residuals from the models in the middle panel. All models detect clear water absorption and a hazy atmosphere. The escaping helium part of the spectrum was excluded and modeled separately.}
  \label{fig:spectrum}
\end{figure*}

\begin{figure*}
\centering
 \includegraphics[width=\textwidth,keepaspectratio]{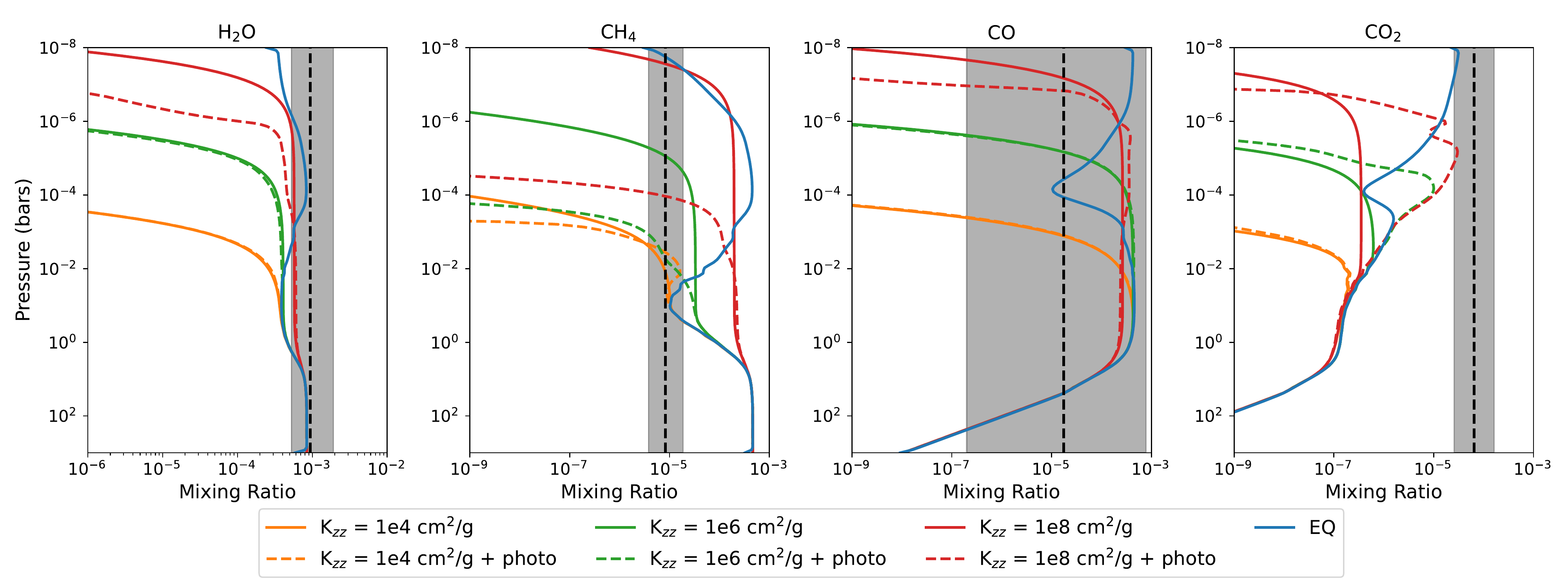}
  \caption{We explored the effects of vertical mixing and photochemistry using VULCAN \citep{tsai_comparative_2021}. The blue line is chemical abundance for H$_2$O, CH$_4$, CO and CO$_2$ at each pressure level from the PHOENIX forward model assuming equilibrium chemistry and solar composition. The orange, green and red lines show the abundance with different $K_{\rm{zz}}$ values. The dashed lines are with photochemistry turned on. The dashed black line and grey shaded regions are the retrieved constraints from ATMO. We see both vertical mixing and photochemistry are effective at removing methane from the atmosphere.}
  \label{fig:diseq}
\end{figure*}

\section{Atmospheric retrievals and modeling}

\subsection{PHOENIX forward model}

PHOENIX is a self-consistent atmospheric forward model assuming Local Thermodynamic Equilibrium (LTE) and equilibrium chemistry \citep{lothringer_extremely_2018}. The transmission spectrum shows distinct haze (optical scattering slope), water, and the lack of strong methane features. A solar composition PHOENIX models fail to fit the spectrum as they largely over predict the methane features (Figure \ref{fig:spectrum}). One way to suppress the methane while preserving water is to increase the metallicity and decrease the C/O ratio. Therefore we generate a PHOENIX forward model grid at 10, 50, and 100 times solar metallicity and C/O ratios of 0.3 and 0.53 with various haze strengths and slopes. The best-fit PHOENIX model ($\chi_{\nu}^2$=1.149 with 1013 degrees of freedom) has a 100 times solar metallicity and 0.3 C/O ratio as shown (red) in the mid-panel of Figure \ref{fig:spectrum}. It under-predicts the 1.4 $\mu m$ water feature while over-predicting the 2.3 $\mu m$ methane feature. This suggests that the equilibrium chemistry model struggles to effectively remove methane without also over-suppressing the water features.

We further explored methane depletion by running models with molecular abundances quenched at 0.1, 1 bar to simulate the effects of vertical mixing. The methane abundance at 0.1 bar is a few orders of magnitude below its maximum abundance, resulting in the most methane depletion if quenching occurs near this level in the atmosphere (Figure \ref{fig:diseq}). We then pick the 0.1 bar as the quench point and were able to reduce the best fit forward model metallicity from 100 to 50 times solar. The 0.1 bar pressure level was selected solely base on examining the methane abundance profile and finding the corresponding pressure level for minimum methane mixing ratio. Only CH$_4$, CO$_2$, CO, and H$_2$O were quenched in our model and quenching approximation does not work for every chemical species \citep{tsai_vulcan_2017}.

We also tested the effects of photochemistry and vertical mixing using the chemical kinetics code VULCAN \citep{tsai_comparative_2021}. We ran VULCAN using the atmosphere structure from the best-fit PHOENIX model with default parameters using the HCNO chemical network with various vertically-constant $K_{\rm{zz}}$ both with and without photochemistry. Abundance profiles of H$_2$O, CH$_4$, CO, and CO$_2$ are shown in Figure \ref{fig:diseq}. Various scenarios with vertical mixing and photochemistry can lead to significant methane-depletion, matching constraints from the retrieval, but determining the specific contribution from each effect at work in HAT-P-18b requires a more detailed retrieval analysis which is beyond of the scope of this paper.

\subsection{ATMO free chemistry retrieval}

To better explore the abundance and significance of the detected chemical species, we also ran a free chemistry retrieval with ATMO \citep{tremblin_fingering_2015, tremblin_cloudless_2016} including temperature, radius, cloud deck pressure, haze strength, haze scattering slope, H$_2$O, CO$_2$, CO, and CH$_4$. We detected water with a volume mixing ratio (VMR) of log(H$_2$O)=-3.03$^{+0.31}_{-0.25}$) and retrieved a methane abundance of log(CH$_4$)=-5.08$^{+0.35}_{-0.34}$. Compared to the PHOENIX forward model, the water abundance is consistent with a $\sim$5 times solar metallicity atmosphere at 1 mbar assuming chemical equilibrium and a solar C/O ratio while methane is about two orders of magnitude lower. We find a CO abundance of log(CO=-4.76$^{+1.65}_{-1.94}$) and a CO$_2$ abundance of log(CO$_2$=-4.19$^{+0.4}_{-0.4}$) are more weakly constrained due to the lack of spectral features at these wavelengths, but their low retrieved abundances are consistent with the forward model predictions since most of the carbon should be in the form of methane. 

To further investigate the presence of methane in the atmosphere we ran another identical retrieval but without CH$_4$. The best-fit retrieval model (BIC=1190.19 and log Bayesian evidence=5570.77) produced a comparable fit to the first full retrieval (BIC=1187.21 and log Bayesian evidence=5574.56) with methane included. The difference of 3.79 in log Bayesian evidence would correspond to moderate evidence \citep{trotta_bayes_2008} supporting the detection of methane. However, the Bayesian information criterion (BIC) favors the retrieval without methane due to one less free fitting parameter. The retrieved VMR of water (H$_2$O)=-3.8$^{+0.17}_{-0.17}$) stays unchanged and CO$_2$ (log(CO$_2$)=-5.72$^{+0.61}_{-0.72}$) remains unconstrained, but the CO (log(CO)=-3.38$^{+0.61}_{-1.02}$) abundance increases. We believe this is due to the overlapping feature from CH$_4$ and CO around 2.3 $\mu m$. Based on the free chemistry retrieval results, we are hesitant to conclude clear detection of methane in the atmosphere of HAT-P-18b. A definitive detection and more stringent limits on CH$_4$, CO and CO$_2$ could be made with NIRSpec G395H, as the molecular signatures of all three carbon-bearing species are considerably stronger at longer wavelengths.

\section{Discussion and Conclusion}

The lack of identifiable methane features in the transmission spectrum (Table \ref{transit_spectrum}) challenges the simple equilibrium chemistry assumption and motivates the exploration of additional physical processes such as vertical mixing and photochemistry \citep{cooper_dynamics_2006, steinrueck_effect_2019, drummond_implications_2020}. Vertical mixing brings up material from deeper in the planet and changes the atmospheric composition probed by the transmission spectrum \citep{fortney_beyond_2020, zahnle_methane_2014}. Depending on the planet's thermal structure and horizontal circulation, mixing could lead to the destruction of methane from hotter interior or dayside of the planet \citep{komacek_vertical_2019}. On the other hand, photochemistry at the top of the atmosphere can break down methane into other hydrocarbon and haze precursors \citep{tsai_comparative_2021}. The detection of a strong haze layer on HAT-P-18b is consistent with the presence of a methane-derived photochemical haze. 


We present the NIRISS/SOSS transmission spectrum of HAT-P-18b, an inflated Saturn-like mass planet with an equilibrium temperature of $\sim$850K. The 0.6 to 2.8$\mu m$ wavelength coverage from SOSS gives us the first look of this planet in the infrared. We detect water, an escaping helium tail, and methane depletion in the hazy atmosphere of HAT-P-18b. The water and helium spectral features are consistent with previous observations \citep{vissapragada_upper_2022, paragas_metastable_2021, tsiaras_population_2018}. The detection of helium tail showcases the unparalleled level of photometric precision from JWST compared to ground-based observation. JWST NIRSpec/G140H (R$\sim$2700) will be the ideal instrument mode for any future study of escape helium.

The methane-depleted atmosphere adds to the growing observational evidence \citep{kreidberg_absence_2019, stevenson_possible_2010, benneke_sub-neptune_2019} of missing methane in sub-1000K planets. This calls for including disequilibrium chemistry processes such as vertical mixing and photochemistry which are effective at delaying the onset of detectable methane features to lower equilibrium temperatures. The fast-growing JWST datasets of exoplanet atmospheres will certainly challenge our simple model assumptions and additional modeling efforts with more physical processes will be needed.

Future follow up observation of HAT-P-18b with JWST NIRSpec/prism covering the 2.8 to 5.3 $\mu m$ wavelength range will be useful at determining how depleted methane is in the atmosphere by measuring the stronger 3 to 4 $\mu m$ methane feature. Additionally, it will better constrain the planet's C/O ratio by simultaneous coverage of CO and CO$_2$ features at 4 and 5 $\mu$m.

\begin{table}
\caption{\textbf{HAT-P-18b transmission spectrum}}
\begin{tabular}{ccc}
\hline\hline & \\
\multicolumn{1}{c}{\shortstack{Wavelength \\ midpoint \\ ($\mu$m)}} & 
\multicolumn{1}{c}{\shortstack{Transit \\ Depth \\ (ppm)}} & 
\multicolumn{1}{c}{\shortstack{Uncertainty \\ (ppm)}}\\[1ex]
\hline 
\setlength\extrarowheight{3pt}
0.6078	&	18908	&	1009    \\
0.6087	&	18903	&	616	\\
0.6096	&	19277	&	598	\\
0.6104	&	19544	&	653	\\
0.6113	&	19468	&	525	\\
\hline 
\multicolumn{3}{p{0.3\textwidth}}{(This table is available in its entirety in machine-readable form.)}
\label{transit_spectrum}
\end{tabular}
\end{table}

\clearpage

\bibliography{references}

\begin{thebibliography}{}
\expandafter\ifx\csname natexlab\endcsname\relax\def\natexlab#1{#1}\fi
\providecommand{\url}[1]{\href{#1}{#1}}
\providecommand{\dodoi}[1]{doi:~\href{http://doi.org/#1}{\nolinkurl{#1}}}
\providecommand{\doeprint}[1]{\href{http://ascl.net/#1}{\nolinkurl{http://ascl.net/#1}}}
\providecommand{\doarXiv}[1]{\href{https://arxiv.org/abs/#1}{\nolinkurl{https://arxiv.org/abs/#1}}}

\bibitem[{Agol {et~al.}(2010)Agol, Cowan, Knutson, Deming, Steffen, Henry, \&
  Charbonneau}]{agol_climate_2010}
Agol, E., Cowan, N.~B., Knutson, H.~A., {et~al.} 2010, The Astrophysical
  Journal, 721, 1861, \dodoi{10.1088/0004-637X/721/2/1861}

\bibitem[{Allart {et~al.}(2019)Allart, Bourrier, Lovis, Ehrenreich, Aceituno,
  Guijarro, Pepe, Sing, Spake, \& Wyttenbach}]{allart_high-resolution_2019}
Allart, R., Bourrier, V., Lovis, C., {et~al.} 2019, Astronomy \& Astrophysics,
  623, A58, \dodoi{10.1051/0004-6361/201834917}

\bibitem[{Benneke {et~al.}(2019)Benneke, Knutson, Lothringer, Crossfield,
  Moses, Morley, Kreidberg, Fulton, Dragomir, Howard, Wong, Désert,
  McCullough, Kempton, Fortney, Gilliland, Deming, \&
  Kammer}]{benneke_sub-neptune_2019}
Benneke, B., Knutson, H.~A., Lothringer, J., {et~al.} 2019, Nature Astronomy,
  3, 813, \dodoi{10.1038/s41550-019-0800-5}

\bibitem[{Cooper \& Showman(2006)}]{cooper_dynamics_2006}
Cooper, C.~S., \& Showman, A.~P. 2006, The Astrophysical Journal, 649, 1048,
  \dodoi{10.1086/506312}

\bibitem[{Dos~Santos {et~al.}(2022)Dos~Santos, Vidotto, Vissapragada, Alam,
  Allart, Bourrier, Kirk, Seidel, \& Ehrenreich}]{dos_santos_p-winds_2022}
Dos~Santos, L.~A., Vidotto, A.~A., Vissapragada, S., {et~al.} 2022, Astronomy
  \& Astrophysics, 659, A62, \dodoi{10.1051/0004-6361/202142038}

\bibitem[{Drummond {et~al.}(2020)Drummond, Hébrard, Mayne, Venot, Ridgway,
  Changeat, Tsai, Manners, Tremblin, Abraham, Sing, \&
  Kohary}]{drummond_implications_2020}
Drummond, B., Hébrard, E., Mayne, N.~J., {et~al.} 2020, Astronomy \&
  Astrophysics, 636, A68, \dodoi{10.1051/0004-6361/201937153}

\bibitem[{Everett {et~al.}(2013)Everett, Howell, Silva, \&
  Szkody}]{everett_spectroscopy_2013}
Everett, M.~E., Howell, S.~B., Silva, D.~R., \& Szkody, P. 2013, The
  Astrophysical Journal, 771, 107, \dodoi{10.1088/0004-637X/771/2/107}

\bibitem[{Foreman-Mackey {et~al.}(2013)Foreman-Mackey, Hogg, Lang, \&
  Goodman}]{foreman-mackey_emcee_2013}
Foreman-Mackey, D., Hogg, D.~W., Lang, D., \& Goodman, J. 2013, Publications of
  the Astronomical Society of the Pacific, 125, 306, \dodoi{10.1086/670067}

\bibitem[{Fortney {et~al.}(2020)Fortney, Visscher, Marley, Hood, Line,
  Thorngren, Freedman, \& Lupu}]{fortney_beyond_2020}
Fortney, J.~J., Visscher, C., Marley, M.~S., {et~al.} 2020, The Astronomical
  Journal, 160, 288, \dodoi{10.3847/1538-3881/abc5bd}

\bibitem[{{Gaia Collaboration} {et~al.}(2018){Gaia Collaboration}, Brown,
  Vallenari, Prusti, de~Bruijne, Babusiaux, Bailer-Jones, Biermann, Evans,
  Eyer, Jansen, Jordi, Klioner, Lammers, Lindegren, Luri, Mignard, Panem,
  Pourbaix, Randich, Sartoretti, Siddiqui, Soubiran, van Leeuwen, Walton,
  Arenou, Bastian, Cropper, Drimmel, Katz, Lattanzi, Bakker, Cacciari,
  Castañeda, Chaoul, Cheek, De~Angeli, Fabricius, Guerra, Holl, Masana,
  Messineo, Mowlavi, Nienartowicz, Panuzzo, Portell, Riello, Seabroke, Tanga,
  Thévenin, Gracia-Abril, Comoretto, Garcia-Reinaldos, Teyssier, Altmann,
  Andrae, Audard, Bellas-Velidis, Benson, Berthier, Blomme, Burgess, Busso,
  Carry, Cellino, Clementini, Clotet, Creevey, Davidson, De~Ridder, Delchambre,
  Dell’Oro, Ducourant, Fernández-Hernández, Fouesneau, Frémat, Galluccio,
  García-Torres, González-Núñez, González-Vidal, Gosset, Guy, Halbwachs,
  Hambly, Harrison, Hernández, Hestroffer, Hodgkin, Hutton, Jasniewicz,
  Jean-Antoine-Piccolo, Jordan, Korn, Krone-Martins, Lanzafame, Lebzelter,
  Löffler, Manteiga, Marrese, Martín-Fleitas, Moitinho, Mora, Muinonen,
  Osinde, Pancino, Pauwels, Petit, Recio-Blanco, Richards, Rimoldini, Robin,
  Sarro, Siopis, Smith, Sozzetti, Süveges, Torra, van Reeven, Abbas,
  Abreu~Aramburu, Accart, Aerts, Altavilla, Álvarez, Alvarez, Alves, Anderson,
  Andrei, Anglada~Varela, Antiche, Antoja, Arcay, Astraatmadja, Bach, Baker,
  Balaguer-Núñez, Balm, Barache, Barata, Barbato, Barblan, Barklem, Barrado,
  Barros, Barstow, Bartholomé~Muñoz, Bassilana, Becciani, Bellazzini,
  Berihuete, Bertone, Bianchi, Bienaymé, Blanco-Cuaresma, Boch, Boeche,
  Bombrun, Borrachero, Bossini, Bouquillon, Bourda, Bragaglia, Bramante,
  Breddels, Bressan, Brouillet, Brüsemeister, Brugaletta, Bucciarelli,
  Burlacu, Busonero, Butkevich, Buzzi, Caffau, Cancelliere, Cannizzaro,
  Cantat-Gaudin, Carballo, Carlucci, Carrasco, Casamiquela, Castellani,
  Castro-Ginard, Charlot, Chemin, Chiavassa, Cocozza, Costigan, Cowell, Crifo,
  Crosta, Crowley, Cuypers†, Dafonte, Damerdji, Dapergolas, David, David,
  de~Laverny, De~Luise, De~March, de~Martino, de~Souza, de~Torres, Debosscher,
  del Pozo, Delbo, Delgado, Delgado, Di~Matteo, Diakite, Diener, Distefano,
  Dolding, Drazinos, Durán, Edvardsson, Enke, Eriksson, Esquej,
  Eynard~Bontemps, Fabre, Fabrizio, Faigler, Falcão, Farràs~Casas, Federici,
  Fedorets, Fernique, Figueras, Filippi, Findeisen, Fonti, Fraile, Fraser,
  Frézouls, Gai, Galleti, Garabato, García-Sedano, Garofalo, Garralda, Gavel,
  Gavras, Gerssen, Geyer, Giacobbe, Gilmore, Girona, Giuffrida, Glass, Gomes,
  Granvik, Gueguen, Guerrier, Guiraud, Gutiérrez-Sánchez, Haigron,
  Hatzidimitriou, Hauser, Haywood, Heiter, Helmi, Heu, Hilger, Hobbs, Hofmann,
  Holland, Huckle, Hypki, Icardi, Janßen, Jevardat~de Fombelle, Jonker,
  Juhász, Julbe, Karampelas, Kewley, Klar, Kochoska, Kohley, Kolenberg,
  Kontizas, Kontizas, Koposov, Kordopatis, Kostrzewa-Rutkowska, Koubsky,
  Lambert, Lanza, Lasne, Lavigne, Le~Fustec, Le~Poncin-Lafitte, Lebreton,
  Leccia, Leclerc, Lecoeur-Taibi, Lenhardt, Leroux, Liao, Licata, Lindstrøm,
  Lister, Livanou, Lobel, López, Managau, Mann, Mantelet, Marchal, Marchant,
  Marconi, Marinoni, Marschalkó, Marshall, Martino, Marton, Mary, Massari,
  Matijevič, Mazeh, McMillan, Messina, Michalik, Millar, Molina, Molinaro,
  Molnár, Montegriffo, Mor, Morbidelli, Morel, Morris, Mulone, Muraveva,
  Musella, Nelemans, Nicastro, Noval, O’Mullane, Ordénovic,
  Ordóñez-Blanco, Osborne, Pagani, Pagano, Pailler, Palacin, Palaversa,
  Panahi, Pawlak, Piersimoni, Pineau, Plachy, Plum, Poggio, Poujoulet, Prša,
  Pulone, Racero, Ragaini, Rambaux, Ramos-Lerate, Regibo, Reylé, Riclet,
  Ripepi, Riva, Rivard, Rixon, Roegiers, Roelens, Romero-Gómez, Rowell, Royer,
  Ruiz-Dern, Sadowski, Sagristà~Sellés, Sahlmann, Salgado, Salguero, Sanna,
  Santana-Ros, Sarasso, Savietto, Schultheis, Sciacca, Segol, Segovia,
  Ségransan, Shih, Siltala, Silva, Smart, Smith, Solano, Solitro, Sordo,
  Soria~Nieto, Souchay, Spagna, Spoto, Stampa, Steele, Steidelmüller,
  Stephenson, Stoev, Suess, Surdej, Szabados, Szegedi-Elek, Tapiador, Taris,
  Tauran, Taylor, Teixeira, Terrett, Teyssandier, Thuillot, Titarenko,
  Torra~Clotet, Turon, Ulla, Utrilla, Uzzi, Vaillant, Valentini, Valette, van
  Elteren, Van~Hemelryck, van Leeuwen, Vaschetto, Vecchiato, Veljanoski, Viala,
  Vicente, Vogt, von Essen, Voss, Votruba, Voutsinas, Walmsley, Weiler, Wertz,
  Wevers, Wyrzykowski, Yoldas, Žerjal, Ziaeepour, Zorec, Zschocke, Zucker,
  Zurbach, \& Zwitter}]{gaia_collaboration_gaia_2018}
{Gaia Collaboration}, Brown, A. G.~A., Vallenari, A., {et~al.} 2018, Astronomy
  \& Astrophysics, 616, A1, \dodoi{10.1051/0004-6361/201833051}

\bibitem[{Guilluy {et~al.}(2020)Guilluy, Andretta, Borsa, Giacobbe, Sozzetti,
  Covino, Bourrier, Fossati, Bonomo, Esposito, Giampapa, Harutyunyan, Rainer,
  Brogi, Bruno, Claudi, Frustagli, Lanza, Mancini, Pino, Poretti, Scandariato,
  Affer, Baffa, Baruffolo, Benatti, Biazzo, Bignamini, Boschin, Carleo,
  Cecconi, Cosentino, Damasso, Desidera, Falcini, Martinez~Fiorenzano, Ghedina,
  González-Álvarez, Guerra, Hernandez, Leto, Maggio, Malavolta, Maldonado,
  Micela, Molinari, Nascimbeni, Pagano, Pedani, Piotto, \&
  Reiners}]{guilluy_gaps_2020}
Guilluy, G., Andretta, V., Borsa, F., {et~al.} 2020, Astronomy \& Astrophysics,
  639, A49, \dodoi{10.1051/0004-6361/202037644}

\bibitem[{Hartman {et~al.}(2012)Hartman, Bakos, Béky, Torres, Latham, Csubry,
  Penev, Shporer, Fulton, Buchhave, Johnson, Howard, Marcy, Fischer, Kovács,
  Noyes, Esquerdo, Everett, Szklenár, Quinn, Bieryla, Knox, Hinz, Sasselov,
  Fűrész, Stefanik, Lázár, Papp, \&
  Sári}]{hartman_hat-p-39bhat-p-41b_2012}
Hartman, J.~D., Bakos, G.~A., Béky, B., {et~al.} 2012, The Astronomical
  Journal, 144, 139, \dodoi{10.1088/0004-6256/144/5/139}

\bibitem[{Kasting {et~al.}(1983)Kasting, Zahnle, \&
  Walker}]{kasting_photochemistry_1983}
Kasting, J.~F., Zahnle, K.~J., \& Walker, J. C.~G. 1983, Precambrian Research,
  20, 121, \dodoi{10.1016/0301-9268(83)90069-4}

\bibitem[{Kempton {et~al.}(2012)Kempton, Zahnle, \&
  Fortney}]{kempton_atmospheric_2012}
Kempton, E. M.-R., Zahnle, K., \& Fortney, J.~J. 2012, The Astrophysical
  Journal, 745, 3, \dodoi{10.1088/0004-637X/745/1/3}

\bibitem[{Kirk {et~al.}(2020)Kirk, Alam, López-Morales, \&
  Zeng}]{kirk_confirmation_2020}
Kirk, J., Alam, M.~K., López-Morales, M., \& Zeng, L. 2020, The Astronomical
  Journal, 159, 115, \dodoi{10.3847/1538-3881/ab6e66}

\bibitem[{Kirk {et~al.}(2017)Kirk, Wheatley, Louden, Doyle, Skillen, McCormac,
  Irwin, \& Karjalainen}]{kirk_rayleigh_2017}
Kirk, J., Wheatley, P.~J., Louden, T., {et~al.} 2017, Monthly Notices of the
  Royal Astronomical Society, 468, 3907, \dodoi{10.1093/mnras/stx752}

\bibitem[{Komacek {et~al.}(2019)Komacek, Showman, \&
  Parmentier}]{komacek_vertical_2019}
Komacek, T.~D., Showman, A.~P., \& Parmentier, V. 2019, The Astrophysical
  Journal, 881, 152, \dodoi{10.3847/1538-4357/ab338b}

\bibitem[{Kreidberg(2015)}]{kreidberg_batman_2015}
Kreidberg, L. 2015, Publications of the Astronomical Society of the Pacific,
  127, 1161, \dodoi{10.1086/683602}

\bibitem[{Kreidberg {et~al.}(2019)Kreidberg, Koll, Morley, Hu, Schaefer,
  Deming, Stevenson, Dittmann, Vanderburg, Berardo, Guo, Stassun, Crossfield,
  Charbonneau, Latham, Loeb, Ricker, Seager, \&
  Vanderspek}]{kreidberg_absence_2019}
Kreidberg, L., Koll, D. D.~B., Morley, C., {et~al.} 2019, Nature, 573, 87,
  \dodoi{10.1038/s41586-019-1497-4}

\bibitem[{Lampón {et~al.}(2021)Lampón, López-Puertas, Sanz-Forcada,
  Sánchez-López, Molaverdikhani, Czesla, Quirrenbach, Pallé, Caballero,
  Henning, Salz, Nortmann, Aceituno, Amado, Bauer, Montes, Nagel, Reiners, \&
  Ribas}]{lampon_modelling_2021}
Lampón, M., López-Puertas, M., Sanz-Forcada, J., {et~al.} 2021, Astronomy \&
  Astrophysics, 647, A129, \dodoi{10.1051/0004-6361/202039417}

\bibitem[{Lothringer {et~al.}(2018)Lothringer, Barman, \&
  Koskinen}]{lothringer_extremely_2018}
Lothringer, J.~D., Barman, T., \& Koskinen, T. 2018, The Astrophysical Journal,
  866, 27, \dodoi{10.3847/1538-4357/aadd9e}

\bibitem[{Magic {et~al.}(2015)Magic, Chiavassa, Collet, \&
  Asplund}]{magic_stagger-grid_2015}
Magic, Z., Chiavassa, A., Collet, R., \& Asplund, M. 2015, Astronomy \&
  Astrophysics, 573, A90, \dodoi{10.1051/0004-6361/201423804}

\bibitem[{Moses {et~al.}(2011)Moses, Visscher, Fortney, Showman, Lewis,
  Griffith, Klippenstein, Shabram, Friedson, Marley, \&
  Freedman}]{moses_disequilibrium_2011}
Moses, J.~I., Visscher, C., Fortney, J.~J., {et~al.} 2011, The Astrophysical
  Journal, 737, 15, \dodoi{10.1088/0004-637X/737/1/15}

\bibitem[{Ngo {et~al.}(2015)Ngo, Knutson, Hinkley, Crepp, Bechter, Batygin,
  Howard, Johnson, Morton, \& Muirhead}]{ngo_friends_2015}
Ngo, H., Knutson, H.~A., Hinkley, S., {et~al.} 2015, The Astrophysical Journal,
  800, 138, \dodoi{10.1088/0004-637X/800/2/138}

\bibitem[{Ngo {et~al.}(2016)Ngo, Knutson, Hinkley, Bryan, Crepp, Batygin,
  Crossfield, Hansen, Howard, Johnson, Mawet, Morton, Muirhead, \&
  Wang}]{ngo_friends_2016}
---. 2016, The Astrophysical Journal, 827, 8, \dodoi{10.3847/0004-637X/827/1/8}

\bibitem[{Nortmann {et~al.}(2018)Nortmann, Pallé, Salz, Sanz-Forcada, Nagel,
  Alonso-Floriano, Czesla, Yan, Chen, Snellen, Zechmeister, Schmitt,
  López-Puertas, Casasayas-Barris, Bauer, Amado, Caballero, Dreizler, Henning,
  Lampón, Montes, Molaverdikhani, Quirrenbach, Reiners, Ribas,
  Sánchez-López, Schneider, \& Osorio}]{nortmann_ground-based_2018}
Nortmann, L., Pallé, E., Salz, M., {et~al.} 2018, Science, 362, 1388,
  \dodoi{10.1126/science.aat5348}

\bibitem[{Oklopčić \& Hirata(2018)}]{oklopcic_new_2018}
Oklopčić, A., \& Hirata, C.~M. 2018, The Astrophysical Journal, 855, L11,
  \dodoi{10.3847/2041-8213/aaada9}

\bibitem[{P.~Loyd {et~al.}(2016)P.~Loyd, France, Youngblood, Schneider, Brown,
  Hu, Linsky, Froning, Redfield, Rugheimer, \& Tian}]{p_loyd_muscles_2016}
P.~Loyd, R.~O., France, K., Youngblood, A., {et~al.} 2016, The Astrophysical
  Journal, 824, 102, \dodoi{10.3847/0004-637X/824/2/102}

\bibitem[{Paragas {et~al.}(2021)Paragas, Vissapragada, Knutson, Oklopčić,
  Chachan, Greklek-McKeon, Dai, Tinyanont, \&
  Vasisht}]{paragas_metastable_2021}
Paragas, K., Vissapragada, S., Knutson, H.~A., {et~al.} 2021, The Astrophysical
  Journal Letters, 909, L10, \dodoi{10.3847/2041-8213/abe706}

\bibitem[{Parker(1958)}]{parker_dynamics_1958}
Parker, E.~N. 1958, The Astrophysical Journal, 128, 664, \dodoi{10.1086/146579}

\bibitem[{Pont {et~al.}(2013)Pont, Sing, Gibson, Aigrain, Henry, \&
  Husnoo}]{pont_prevalence_2013}
Pont, F., Sing, D.~K., Gibson, N.~P., {et~al.} 2013, Monthly Notices of the
  Royal Astronomical Society, 432, 2917, \dodoi{10.1093/mnras/stt651}

\bibitem[{Pontoppidan {et~al.}(2022)Pontoppidan, Blome, Braun, Brown,
  Carruthers, Coe, DePasquale, Espinoza, Marin, Gordon, Henry, Hustak, James,
  Koekemoer, LaMassa, Law, Lockwood, Moro-Martin, Mullally, Pagan, Player,
  Proffitt, Pulliam, Ramsay, Ravindranath, Reid, Robberto, Sabbi, \&
  Ubeda}]{pontoppidan_jwst_2022}
Pontoppidan, K., Blome, C., Braun, H., {et~al.} 2022, The {JWST} {Early}
  {Release} {Observations},  arXiv.
\newblock \url{http://arxiv.org/abs/2207.13067}

\bibitem[{Seager \& Sasselov(2000)}]{seager_theoretical_2000}
Seager, S., \& Sasselov, D.~D. 2000, The Astrophysical Journal, 537, 916,
  \dodoi{10.1086/309088}

\bibitem[{Sing {et~al.}(2011)Sing, Pont, Aigrain, Charbonneau, Desert, Gibson,
  Gilliland, Hayek, Henry, Knutson, Etangs, Mazeh, \&
  Shporer}]{sing_hubble_2011}
Sing, D.~K., Pont, F., Aigrain, S., {et~al.} 2011, Monthly Notices of the Royal
  Astronomical Society, 416, 1443, \dodoi{10.1111/j.1365-2966.2011.19142.x}

\bibitem[{Spake {et~al.}(2021)Spake, Oklopčić, \&
  Hillenbrand}]{spake_post-transit_2021}
Spake, J.~J., Oklopčić, A., \& Hillenbrand, L.~A. 2021, The Astronomical
  Journal, 162, 284, \dodoi{10.3847/1538-3881/ac178a}

\bibitem[{Steinrueck {et~al.}(2019)Steinrueck, Parmentier, Showman, Lothringer,
  \& Lupu}]{steinrueck_effect_2019}
Steinrueck, M.~E., Parmentier, V., Showman, A.~P., Lothringer, J.~D., \& Lupu,
  R.~E. 2019, The Astrophysical Journal, 880, 14,
  \dodoi{10.3847/1538-4357/ab2598}

\bibitem[{Stevenson {et~al.}(2010)Stevenson, Harrington, Nymeyer, Madhusudhan,
  Seager, Bowman, Hardy, Deming, Rauscher, \& Lust}]{stevenson_possible_2010}
Stevenson, K.~B., Harrington, J., Nymeyer, S., {et~al.} 2010, Nature, 464,
  1161, \dodoi{10.1038/nature09013}

\bibitem[{The JWST Transiting Exoplanet Community Early Release Science~Team
  {et~al.}(2022)The JWST Transiting Exoplanet Community Early Release
  Science~Team, Ahrer, Alderson, Batalha, Batalha, Bean, Beatty, Bell, Benneke,
  Berta-Thompson, Carter, Crossfield, Espinoza, Feinstein, Fortney, Gibson,
  Goyal, Kempton, Kirk, Kreidberg, López-Morales, Line, Lothringer, Moran,
  Mukherjee, Ohno, Parmentier, Piaulet, Rustamkulov, Schlawin, Sing, Stevenson,
  Wakeford, Allen, Birkmann, Brande, Crouzet, Cubillos, Damiano, Désert, Gao,
  Harrington, Hu, Kendrew, Knutson, Lagage, Leconte, Lendl, MacDonald, May,
  Miguel, Molaverdikhani, Moses, Murray, Nehring, Nikolov, de~la Roche, Radica,
  Roy, Stassun, Taylor, Waalkes, Wachiraphan, Welbanks, Wheatley, Aggarwal,
  Alam, Banerjee, Barstow, Blecic, Casewell, Changeat, Chubb, Colón, Coulombe,
  Daylan, de~Val-Borro, Decin, Santos, Flagg, France, Fu, Muñoz, Gizis,
  Glidden, Grant, Heng, Henning, Hong, Inglis, Iro, Kataria, Komacek, Krick,
  Lee, Lewis, Lillo-Box, Lustig-Yaeger, Mancini, Mandell, Mansfield, Marley,
  Mikal-Evans, Morello, Nixon, Ceballos, Piette, Powell, Rackham, Ramos-Rosado,
  Rauscher, Redfield, Rogers, Roman, Roudier, Scarsdale, Shkolnik, Southworth,
  Spake, Steinrueck, Tan, Teske, Tremblin, Tsai, Tucker, Turner, Valenti,
  Venot, Waldmann, Wallack, Zhang, \&
  Zieba}]{the_jwst_transiting_exoplanet_community_early_release_science_team_identification_2022}
The JWST Transiting Exoplanet Community Early Release Science~Team, N., Ahrer,
  E.-M., Alderson, L., {et~al.} 2022, Identification of carbon dioxide in an
  exoplanet atmosphere,  arXiv.
\newblock \url{http://arxiv.org/abs/2208.11692}

\bibitem[{Thompson {et~al.}(2022)Thompson, Krissansen-Totton, Wogan, Telus, \&
  Fortney}]{thompson_case_2022}
Thompson, M.~A., Krissansen-Totton, J., Wogan, N., Telus, M., \& Fortney, J.~J.
  2022, Proceedings of the National Academy of Sciences, 119, e2117933119,
  \dodoi{10.1073/pnas.2117933119}

\bibitem[{Tremblin {et~al.}(2016)Tremblin, Amundsen, Chabrier, Baraffe,
  Drummond, Hinkley, Mourier, \& Venot}]{tremblin_cloudless_2016}
Tremblin, P., Amundsen, D.~S., Chabrier, G., {et~al.} 2016, The Astrophysical
  Journal, 817, L19, \dodoi{10.3847/2041-8205/817/2/L19}

\bibitem[{Tremblin {et~al.}(2015)Tremblin, Amundsen, Mourier, Baraffe,
  Chabrier, Drummond, Homeier, \& Venot}]{tremblin_fingering_2015}
Tremblin, P., Amundsen, D.~S., Mourier, P., {et~al.} 2015, The Astrophysical
  Journal, 804, L17, \dodoi{10.1088/2041-8205/804/1/L17}

\bibitem[{Trotta(2008)}]{trotta_bayes_2008}
Trotta, R. 2008, Contemporary Physics, 49, 71,
  \dodoi{10.1080/00107510802066753}

\bibitem[{Tsai {et~al.}(2017)Tsai, Lyons, Grosheintz, Rimmer, Kitzmann, \&
  Heng}]{tsai_vulcan_2017}
Tsai, S.-M., Lyons, J.~R., Grosheintz, L., {et~al.} 2017, The Astrophysical
  Journal Supplement Series, 228, 20, \dodoi{10.3847/1538-4365/228/2/20}

\bibitem[{Tsai {et~al.}(2021)Tsai, Malik, Kitzmann, Lyons, Fateev, Lee, \&
  Heng}]{tsai_comparative_2021}
Tsai, S.-M., Malik, M., Kitzmann, D., {et~al.} 2021, The Astrophysical Journal,
  923, 264, \dodoi{10.3847/1538-4357/ac29bc}

\bibitem[{Tsiaras {et~al.}(2018)Tsiaras, Waldmann, Zingales, Rocchetto,
  Morello, Damiano, Karpouzas, Tinetti, McKemmish, Tennyson, \&
  Yurchenko}]{tsiaras_population_2018}
Tsiaras, A., Waldmann, I.~P., Zingales, T., {et~al.} 2018, The Astronomical
  Journal, 155, 156, \dodoi{10.3847/1538-3881/aaaf75}

\bibitem[{Vissapragada {et~al.}(2022)Vissapragada, Knutson, Greklek-McKeon,
  Oklopcic, Dai, Santos, Jovanovic, Mawet, Millar-Blanchaer, Paragas, Spake, \&
  Vasisht}]{vissapragada_upper_2022}
Vissapragada, S., Knutson, H.~A., Greklek-McKeon, M., {et~al.} 2022, The
  {Upper} {Edge} of the {Neptune} {Desert} {Is} {Stable} {Against}
  {Photoevaporation},  arXiv.
\newblock \url{http://arxiv.org/abs/2204.11865}

\bibitem[{Youngblood {et~al.}(2016)Youngblood, France, Loyd, Linsky, Redfield,
  Schneider, Wood, Brown, Froning, Miguel, Rugheimer, \&
  Walkowicz}]{youngblood_muscles_2016}
Youngblood, A., France, K., Loyd, R. O.~P., {et~al.} 2016, The Astrophysical
  Journal, 824, 101, \dodoi{10.3847/0004-637X/824/2/101}

\bibitem[{Zahnle \& Marley(2014)}]{zahnle_methane_2014}
Zahnle, K.~J., \& Marley, M.~S. 2014, The Astrophysical Journal, 19

\end{thebibliography}

\end{document}